\documentclass[11pt]{article}

\usepackage{epsfig,amssymb}

\newtheorem{theorem}{Theorem}

\newcommand{\mod}{\ \rm mod \,}

\begin{document}

\title{Simple Sets of Measurements for Universal Quantum Computation and
Graph State Preparation}

\author{Yasuhiro Takahashi\footnote{NTT Communication Science
Laboratories, NTT Corporation}}

\date{}

\maketitle

\begin{abstract}
 We consider the problem of minimizing resources required for universal
 quantum computation using only projective measurements. The resources
 we focus on are observables, which describe projective measurements,
 and ancillary qubits. We show that the set of observables $\{Z\otimes
 X, (\cos\theta) X + (\sin\theta) Y \ {\rm all} \ \theta \in [0,2\pi)\}$
 with one ancillary qubit is universal for quantum computation. The set
 is simpler than a previous one in the sense that one-qubit projective
 measurements described by the observables in the set are ones only in
 the $(X,Y)$ plane of the Bloch sphere. The proof of the universality
 immediately implies a simple set of observables that is approximately
 universal for quantum computation. Moreover, the proof implies a simple
 set of observables for preparing graph states efficiently.
\end{abstract}

\section{Introduction}

In 2001, Raussendorf and Briegel proposed a new model for quantum
computation, which is called cluster state
computation \cite{Raussendorf}. Later, in 2003, based on the idea of
Gottesman and Chuang \cite{Gottesman}, Nielsen proposed a new model,
which is called teleportation-based quantum computation
\cite{Nielsen1}. In contrast to conventional models, such as the quantum
circuit model \cite{Nielsen2}, these new models use only projective
measurements for universal quantum computation and thus suggest a new
way of realizing a quantum computer. Minimizing the resources required
for universal quantum computation is important for realizing a quantum
computer based on these new models.

We consider the problem under the assumption that, as in the
teleportation-based quantum computation and its simplified
version \cite{Perdrix1,Childs}, we can use only projective measurements
and do not have initial cluster states. The resources we focus on are
observables, which describe projective measurements, and ancillary
qubits. There have been many studies in this
direction \cite{Perdrix1,Childs,Leung,Perdrix2,Perdrix3}. In particular,
in 2005, Jorrand and Perdrix showed that the set of observables
$$\{Z\otimes X,Z,(\cos\theta)X + (\sin\theta)Y \ {\rm all} \
\theta\in[0,2\pi)\}$$
with one ancillary qubit is universal for quantum
computation \cite{Perdrix2}, where $X$, $Y$, and $Z$ are Pauli
matrices. It has not been known whether a simpler universal set of
observables can be constructed without increasing the number of
ancillary qubits.

In this paper, we show that the set of observables
$${\cal S}_1=\{Z\otimes X,(\cos\theta)X + (\sin\theta)Y \ {\rm all} \
\theta\in [0,2\pi)\}$$
with one ancillary qubit is universal for quantum computation. The set
is simpler than Jorrand and Perdrix's \cite{Perdrix2} in the sense that
one-qubit projective measurements described by the observables in ${\cal
S}_1$ are ones only in the $(X,Y)$ plane of the Bloch sphere. In the
proof of the universality, the key idea is to use $Y$-measurements
appropriately in place of other one-qubit projective measurements, such
as $X$- and $Z$-measurements. In contrast to Jorrand and Perdrix's
proof \cite{Perdrix2}, our proof connects a simple universal set of
observables with a simple approximate universal one. More precisely, our
proof immediately implies the best known result for the approximate
universality by Perdrix \cite{Perdrix3} that a set of two one-qubit
observables and one two-qubit observable with one ancillary qubit is
approximately universal for quantum computation. For example, our proof
immediately implies that the set of observables
$${\cal S}_2=\{Z\otimes X,Y,(X+Y)/\sqrt{2}\}$$
with one ancillary qubit is approximately universal for quantum
computation.

We also consider the problem of minimizing the resources required for
preparing graph states efficiently. It is important to investigate this
problem since graph states play a key role in quantum information
processing \cite{Hein}. In 2006, H\o yer et al. showed that, for any
graph $G=(V,E)$, some signed graph state $|G\rangle$ can be prepared by
a quantum circuit consisting of one-qubit and two-qubit projective
measurements with size $O(|V|+|E|)$, depth $O(|E|)$, and one ancillary
qubit \cite{Hoyer}. The circuit uses the set of observables in
\cite{Perdrix1}. Even if its improved version in \cite{Perdrix3} is used
in the circuit, two one-qubit observables and one two-qubit observable
are required.

Using the proof of the universality of ${\cal S}_1$, we show that the
set of observables
$${\cal S}_3=\{Z\otimes X,Y\}$$
with one ancillary qubit is sufficient for preparing graph states
efficiently. More precisely, we show that, for any graph $G=(V,E)$, the
(exact) graph state $|G\rangle$ can be prepared by a quantum circuit
consisting of one-qubit and two-qubit projective measurements described
by the observables in ${\cal S}_3$ with size and depth $O(|V|+|E|)$ and
one ancillary qubit. The depth is $O(|E|)$ for the graphs in which we
are interested. Though the usual method for preparing graph states
performs controlled-$Z$ operations repeatedly, it is difficult to do so
since ${\cal S}_3$ has only $Z\otimes X$ and $Y$. The key idea is to
perform operations similar to controlled-$Z$ operations and to remove
the side effects of the similar operations by using $Y$-measurements.

\section{Preliminaries}

\subsection{Simulation of a unitary operation by measurements}

Frequently used observables are Pauli matrices $X$, $Y$, and $Z$ defined
by
$$\left(
\begin{array}{cc}
0 & 1\\
1 & 0
\end{array}
\right),
\left(
\begin{array}{cc}
0 & -i\\
i & 0
\end{array}
\right),
\left(
\begin{array}{cccc}
1 & 0 \\
0 & -1
\end{array}
\right),$$
respectively. They describe the one-qubit projective measurements in the
basis $\{|+_0\rangle,|-_0\rangle\}$,
$\{|+_{\frac{\pi}{2}}\rangle,|-_{\frac{\pi}{2}}\rangle\}$, and
$\{|0\rangle,|1\rangle\}$, respectively, where
$$|+_{\theta}\rangle = \frac{|0\rangle + e^{i\theta}
|1\rangle}{\sqrt{2}},\ 
|-_{\theta}\rangle = \frac{|0\rangle - e^{i\theta}
|1\rangle}{\sqrt{2}}$$
for any $\theta \in [0,2\pi)$. Each basis corresponds to the classical
outcomes $1$ and $-1$, respectively. We denote $|\pm_0\rangle$ as
$|\pm\rangle$. Pauli matrices also denote unitary operations and we use
$\sigma_x$, $\sigma_y$, and $\sigma_z$ in the case. In general, the
observable $(\cos\theta)X + (\sin\theta)Y$ for any $\theta \in [0,2\pi)$
describes the one-qubit projective measurement in the basis
$\{|+_{\theta}\rangle,|-_{\theta}\rangle\}$, where the corresponding
classical outcomes are 1 and $-1$, respectively. This is a projective
measurement in the $(X,Y)$ plane of the Bloch sphere. We also consider
two-qubit observables such as $Z\otimes X$, where $\otimes$ denotes the
tensor product. The projective measurement described by $Z\otimes X$ has
only two possible classical outcomes $1$ and $-1$. It consists of two
projections: one is on the space spanned by $|0\rangle|+\rangle$ and
$|1\rangle|-\rangle$ and the other is on the space spanned by
$|0\rangle|-\rangle$ and $|1\rangle|+\rangle$.

Let $\cal S$ be a set of observables and $U$ be a unitary operation. The
simulation of $U$ by using projective measurements described by the
observables in $\cal S$ is decomposed into the following
steps \cite{Perdrix3}:
\begin{enumerate}
\item Simulation step: $\sigma U$ is probabilistically implemented by
      using projective measurements described by the observables in
      $\cal S$, where $\sigma$ is $\sigma_x$, $\sigma_y$, $\sigma_z$, or
      an identity operation $I$ when $U$ is on one qubit, and is known
      by the classical outcomes of the measurements. When $U$ is on
      multiple qubits, $\sigma$ is allowed to be a tensor product of
      these operations.

\item Correction step: If $\sigma U$ is implemented in the simulation
      step where $\sigma\neq I$, $\sigma$ is implemented by using
      projective measurements described by the observables in $\cal S$
      to obtain $\sigma\sigma U=U$.
\end{enumerate}

\subsection{Universality of a set of observables}

In the quantum circuit model, a set of gates is said to be universal for
quantum computation if any unitary operation can be implemented exactly
by a quantum circuit consisting only of gates in the set. The
approximate universality of a set of gates is defined
similarly \cite{Nielsen2}. It is known that the set of all one-qubit
gates and controlled-$Z$ gate $\Lambda Z$ are universal for quantum
computation \cite{Barenco} and that the set of Hadamard gate $H$, $\pi/8$
gate $Z(\pi/4)$, and $\Lambda Z$ are approximately universal for quantum
computation \cite{Boykin}, where $H$, $Z(\theta)$, and $\Lambda Z$ are
defined by
$$
\frac{1}{\sqrt{2}}
\left(
\begin{array}{cc}
1 & 1\\
1 & -1
\end{array}
\right),
\left(
\begin{array}{cccc}
1 & 0 \\
0 & e^{i\theta}
\end{array}
\right),
\left(
\begin{array}{cccc}
1 & 0 & 0 & 0 \\
0 & 1 & 0 & 0 \\
0 & 0 & 1 & 0 \\
0 & 0 & 0 & -1
\end{array}
\right),
$$
respectively, for any $\theta\in [0,2\pi)$. Moreover, it is known that
$J(\theta)=HZ(\theta)$ generates any one-qubit
gate \cite{Perdrix2,Danos}. A set of observables $\cal S$ is said to be
universal (resp. approximately universal) for quantum computation if
there exists a universal (resp. approximately universal) set of gates
such that any gate (that is, unitary operation) in the set can be
simulated by using projective measurements described by the observables
in $\cal S$.

\section{Our Universal Set of Observables}

\begin{figure}[t]
\centerline{\psfig{file=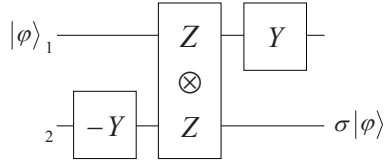,scale=0.35}}
\vspace{-4.2cm}
\caption{The state transfer based on $Y$-measurements.}
\label{figure0}
\end{figure}

The simulation step of $J(\theta)$ in \cite{Perdrix2} is based
on the state transfer, which is a simplified version of quantum
teleportation and uses $X$- and $Z\otimes Z$-measurements. For example,
it implies a simulation step of $J(0)=H$ using $Z$-, $X$-, and $Z\otimes
X$-measurements. To simplify this, our idea is to use the state transfer
based on $Y$-measurements depicted in Fig. \ref{figure0}, which 
transfers the input state $|\varphi\rangle$ from qubit 1 to qubit 2 (up
to Pauli operations). As in \cite{Perdrix2}, this implies a
simulation step of a unitary operation using projective measurements
depending on the operation. For example, we can obtain a simulation step
of $H$ by replacing $-Y$ and $Z\otimes Z$ with $H(-Y)H^{\dagger}=Y$ and
$Z\otimes (HZH^{\dagger})=Z\otimes X$, respectively. The simulation step
is simpler than the previous one since it uses only $Y$- and $Z\otimes
X$-measurements.

On the basis of the idea, we show the following theorem:
\begin{theorem}
The set of observables
$${\cal S}_1=\{Z\otimes X,(\cos\theta)X + (\sin\theta)Y \ {\rm all} \
\theta\in [0,2\pi)\}$$
with one ancillary qubit is universal for quantum computation.
\end{theorem}
{\it Proof.}
The set of gates
$$\{(P^{-1}\otimes HP^{-1})\Lambda Z(I\otimes H),J(\theta) \ {\rm all} \
 \theta\in[0,2\pi)\}$$
is universal for quantum computation, where $P=Z(\pi/2)$ (and thus
 $P^{-1}=\sigma_z P$). This is because $J(\theta)$ generates any
 one-qubit gate and $\{\Lambda Z,J(\theta) \ {\rm all} \
 \theta\in[0,2\pi)\}$ is universal for quantum
 computation \cite{Perdrix2,Danos}. Thus, to show the theorem, it
 suffices to simulate any gate in the above set by projective
 measurements described by the observables in ${\cal S}_1$.

\begin{figure}[t]
\centerline{\psfig{file=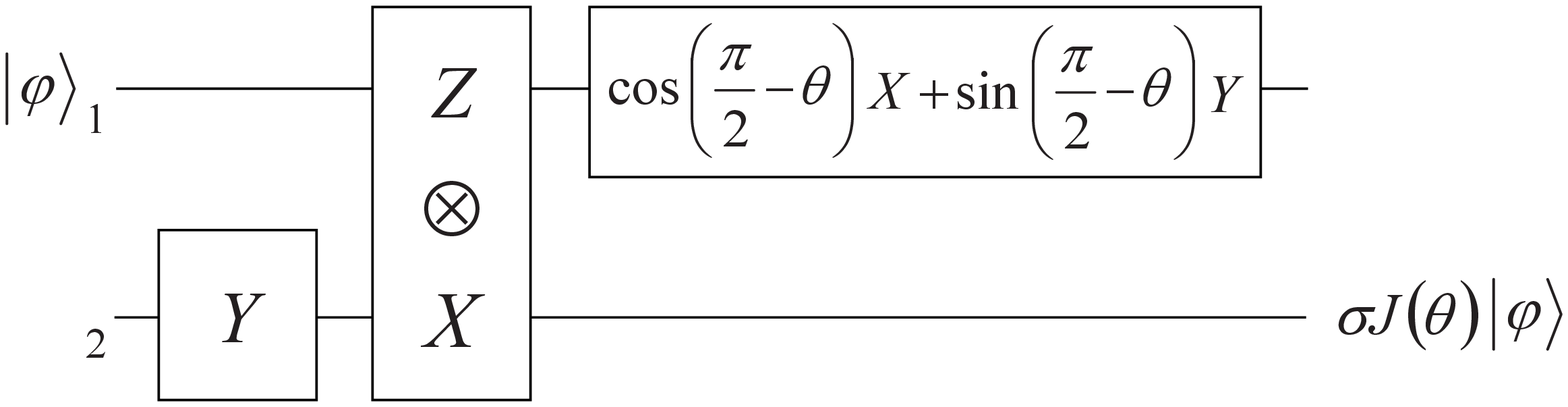,scale=0.35}}
\vspace{-4.2cm}
\caption{The simulation step of $J(\theta)$.}
\label{figure1}
\end{figure}

To give the simulation step of $J(\theta)$, we consider the procedure
 depicted in Fig. \ref{figure1}, which is obtained by replacing $-Y$,
 $Z\otimes Z$, and $Y$ in Fig. \ref{figure0} with $H(-Y)H^{\dagger}=Y$,
 $(Z(\theta)^{\dagger}ZZ(\theta))\otimes (HZH^{\dagger})=Z\otimes X$,
 and $Z(\theta)^{\dagger}YZ(\theta) = \cos(\pi/2-\theta)X +
 \sin(\pi/2-\theta)Y$, respectively. Let
 $|\varphi\rangle=\alpha|0\rangle+\beta|1\rangle$ and $s_1,s_2,s_3\in
 \{1,-1\}$ be the classical outcomes of the measurements $Y^{(2)}$,
 $Z^{(1)}\otimes X^{(2)}$, and
 $(\cos(\pi/2-\theta)X+\sin(\pi/2-\theta)Y)^{(1)}$, respectively. The
 first measurement transforms the input state into
$$(I\otimes
 \sigma^{\frac{1-s_1}{2}}_z)(\alpha|0\rangle+\beta|1\rangle)
 |+_{\frac{\pi}{2}}\rangle.$$
The second measurement transforms the state into
$$(\sigma^{\frac{1-s_1s_2}{2}}_z\otimes
 \sigma^{\frac{1-s_2}{2}}_z)(\alpha|0\rangle|+\rangle -
 i\beta|1\rangle|-\rangle).$$
The third measurement transforms it into
$$(\sigma^{\frac{1-s_3}{2}}_z\otimes \sigma^{\frac{1-s_2}{2}}_z
 \sigma^{\frac{1+s_1s_2s_3}{2}}_x)|+_{\frac{\pi}{2}-\theta}\rangle
 (\alpha|+\rangle+  e^{i\theta}\beta |-\rangle),$$
which is the desired output state since
 $J(\theta)|\varphi\rangle=\alpha|+\rangle+  e^{i\theta}\beta
 |-\rangle$. Thus, the procedure depicted in Fig. \ref{figure1} is a
 simulation step of $J(\theta)$, where $\sigma=I,\ \sigma_x,\ \sigma_z,$
 or $\sigma_z\sigma_x$ $(=\sigma_y$ up to a global phase). It can be
 shown that each $\sigma$ occurs with the same probability, 1/4.

\begin{figure}[t]
\centerline{\psfig{file=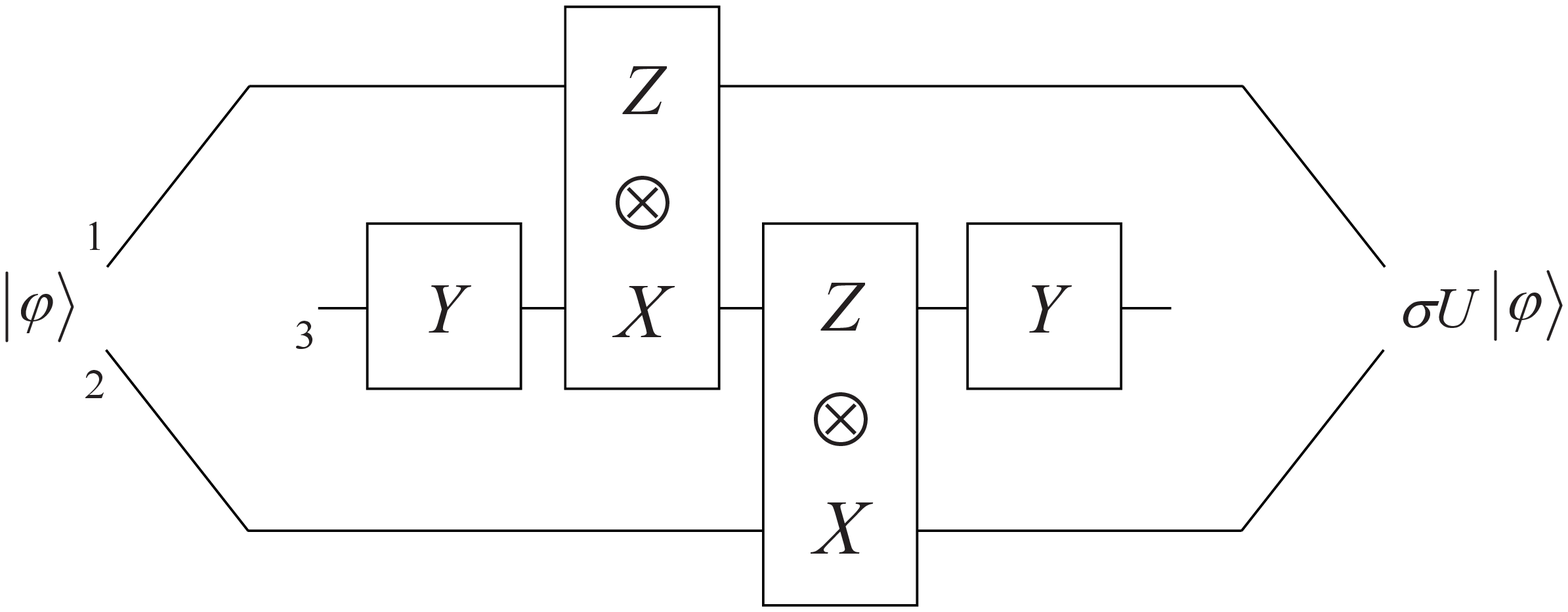,scale=0.35}}
\vspace{-3cm}
\caption{The simulation step of $U=(P^{-1}\otimes HP^{-1})\Lambda
 Z(I\otimes H)$.}
\label{figure2}
\end{figure}

To give the simulation step of $(P^{-1}\otimes HP^{-1})\Lambda
 Z(I\otimes H)$, we consider the procedure depicted in
 Fig. \ref{figure2}. Let $|\varphi\rangle = \alpha|00\rangle +
 \beta|01\rangle + \gamma|10\rangle + \delta|11\rangle$ and
 $s_1,s_2,s_3,s_4\in \{1,-1\}$ be the classical outcomes of the
 measurements $Y^{(3)}$ (the left one), $Z^{(1)}\otimes X^{(3)}$,
 $Z^{(3)}\otimes X^{(2)}$, and $Y^{(3)}$ (the right one),
 respectively. The first measurement transforms the input state into
$$(I\otimes I \otimes \sigma^{\frac{1-s_1}{2}}_z)(\alpha|00\rangle +
 \beta|01\rangle + \gamma|10\rangle +
 \delta|11\rangle)|+_{\frac{\pi}{2}}\rangle.$$
The second measurement transforms the state into
$$(\sigma^{\frac{1-s_1s_2}{2}}_z\otimes I \otimes
 \sigma^{\frac{1-s_2}{2}}_z)(\alpha|0\rangle|0\rangle|+\rangle +
 \beta|0\rangle|1\rangle|+\rangle
-i\gamma|1\rangle|0\rangle|-\rangle
-i\delta|1\rangle|1\rangle|-\rangle).$$
The third measurement transforms the state into
\begin{eqnarray*}
(\sigma^{\frac{1-s_1s_2s_3}{2}}_z \otimes
   \sigma^{\frac{1-s_2}{2}}_x     \otimes
   \sigma^{\frac{1-s_3}{2}}_x)
&&\hspace{-.2cm} (\alpha|0\rangle \frac{|+\rangle|0\rangle +
   |-\rangle|1\rangle}{\sqrt{2}}
+  \beta|0\rangle \frac{|+\rangle|0\rangle -
   |-\rangle|1\rangle}{\sqrt{2}}\\
&&\hspace{-.4cm} -i \gamma|1\rangle  \frac{|+\rangle|0\rangle -
   |-\rangle|1\rangle}{\sqrt{2}}
-i \delta|1\rangle  \frac{|+\rangle|0\rangle +
   |-\rangle|1\rangle}{\sqrt{2}}).
\end{eqnarray*}
The fourth measurement transforms it into
\begin{eqnarray*}
(\sigma^{\frac{1-s_1s_2s_3}{2}}_z \otimes
   \sigma^{\frac{1-s_2s_3s_4}{2}}_x     \otimes
   \sigma^{\frac{1-s_4}{2}}_z)
&&\hspace{-.2cm} (\alpha|0\rangle \frac{|+\rangle -i
   |-\rangle}{\sqrt{2}}
+  \beta|0\rangle \frac{|+\rangle +i
   |-\rangle}{\sqrt{2}}\\
&&\hspace{-.4cm} -i \gamma|1\rangle  \frac{|+\rangle +i
   |-\rangle}{\sqrt{2}}
-i \delta|1\rangle  \frac{|+\rangle -i
   |-\rangle}{\sqrt{2}})|+_{\frac{\pi}{2}}\rangle,
\end{eqnarray*}
which is the desired output state since
\begin{eqnarray*}
(P^{-1}\otimes HP^{-1})\Lambda Z(I\otimes H)|\varphi\rangle
&=& \alpha|0\rangle \frac{|+\rangle -i
   |-\rangle}{\sqrt{2}} 
+  \beta|0\rangle \frac{|+\rangle +i
   |-\rangle}{\sqrt{2}}\\
&& -i \gamma|1\rangle  \frac{|+\rangle +i
   |-\rangle}{\sqrt{2}}
-i \delta|1\rangle  \frac{|+\rangle -i
   |-\rangle}{\sqrt{2}}.
\end{eqnarray*}
Thus, the procedure depicted in Fig. \ref{figure2} is a simulation step
 of $(P^{-1}\otimes HP^{-1})\Lambda Z(I\otimes H)$, where
 $\sigma=I\otimes I,\ I\otimes\sigma_x,\ \sigma_z\otimes I,$ or
 $\sigma_z\otimes\sigma_x$. It can be shown that each $\sigma$ occurs
 with the same probability, 1/4.

\begin{figure}[t]
\centerline{\psfig{file=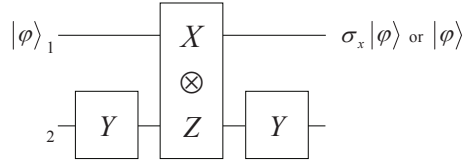,scale=0.35}}
\vspace{-4.2cm}
\caption{The implementation of $\sigma_x$ in the correction step.}
\label{figure3}
\end{figure}

 To implement $\sigma_x$, we consider the procedure depicted in
 Fig. \ref{figure3}. Let
 $|\varphi\rangle=\alpha|0\rangle+\beta|1\rangle$ and $s_1,s_2,s_3\in
 \{1,-1\}$ be the classical outcomes of the measurements $Y^{(2)}$ (the
 left one), $X^{(1)}\otimes Z^{(2)}$, and $Y^{(2)}$ (the right one),
 respectively. The first measurement transforms the input state into
$$(I\otimes \sigma^{\frac{1-s_1}{2}}_z)
 (\alpha|0\rangle+\beta|1\rangle)|+_{\frac{\pi}{2}}\rangle.$$
The second measurement transforms the state into
$$(I \otimes \sigma^{\frac{1-s_1}{2}}_z
 \sigma^{\frac{1-s_2}{2}}_y)
\frac{1}{\sqrt{2}}((\alpha|0\rangle+\beta|1\rangle)|+_{\frac{\pi}{2}}\rangle+
 (\alpha|1\rangle+\beta|0\rangle)|-_{\frac{\pi}{2}}\rangle).$$
The third measurement transforms the state into
$$(\sigma^{\frac{1-s_1s_3}{2}}_x\otimes \sigma^{\frac{1-s_3}{2}}_z)
 (\alpha|0\rangle+\beta|1\rangle)|+_{\frac{\pi}{2}}\rangle.$$
It can be shown that $\sigma_x$ is implemented with the probability
 1/2.

\begin{figure}[t]
\centerline{\psfig{file=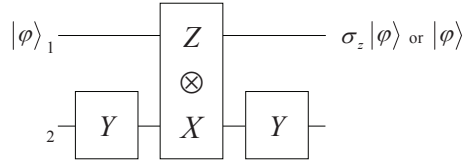,scale=0.35}}
\vspace{-4.2cm}
\caption{The implementation of $\sigma_z$ in the correction step.}
\label{figure4}
\end{figure}

To implement $\sigma_z$, we consider the procedure depicted in
 Fig. \ref{figure4}. Let
 $|\varphi\rangle=\alpha|0\rangle+\beta|1\rangle$ and $s_1,s_2,s_3\in
 \{1,-1\}$ be the classical outcomes of the measurements $Y^{(2)}$ (the
 left one), $Z^{(1)}\otimes X^{(2)}$, and $Y^{(2)}$ (the right one),
 respectively. The first measurement transforms the input state into
$$(I\otimes
 \sigma^{\frac{1-s_1}{2}}_z)(\alpha|0\rangle+\beta|1\rangle)
 |+_{\frac{\pi}{2}}\rangle.$$
The second measurement transforms the state into
$$(\sigma^{\frac{1-s_1s_2}{2}}_z\otimes
 \sigma^{\frac{1-s_2}{2}}_z)(\alpha|0\rangle|+\rangle -
 i\beta|1\rangle|-\rangle).$$
The third measurement transforms the state into
$$(\sigma^{\frac{1-s_1s_3}{2}}_z\otimes \sigma^{\frac{1-s_3}{2}}_z)
 (\alpha|0\rangle+\beta|1\rangle)|+_{\frac{\pi}{2}}\rangle.$$
It can be shown that $\sigma_z$ is implemented with the probability
 1/2.

In the correction step, we repeat the procedures until the desired gate
 $(\sigma_x$ or $\sigma_z)$ is implemented (as in \cite{Perdrix3}). The
 gate $\sigma_y$ is implemented by combining the procedures. Thus, any
 gate in the set described at the beginning of the proof can be
 simulated by projective measurements described by the observables in
 ${\cal S}_1$. \hfill $\Box$\\

The proof of Theorem 1 immediately implies Perdrix's
result \cite{Perdrix3}:
\begin{theorem}
The set of observables
$${\cal S}_2=\{Z\otimes X,Y,(X+Y)/\sqrt{2}\}$$
with one ancillary qubit is approximately universal for quantum
 computation.
\end{theorem}
{\it Proof.}
The set of gates $\{H,J(\pi/4),(P^{-1}\otimes HP^{-1})\Lambda Z(I\otimes
 H)\}$ is approximately universal for quantum computation. This is
 because $\{H,Z(\pi/4),\Lambda Z\}$ is approximately universal for
 quantum computation \cite{Nielsen2,Boykin}, $H^2=I$, and
 $(HJ(\pi/4))^2=P$. On the basis of the set of gates, it is easy to show
 the theorem since the simulation steps of the gates and the correction
 steps in the proof of Theorem 1 use projective measurements described
 by the observables only in ${\cal S}_2$.\hfill $\Box$\\

We can also immediately imply other approximately universal sets of
observables using other approximately universal sets of gates
\cite{Shi}.

\section{Efficient Graph State Preparation}

\begin{figure}[t]
\centerline{\psfig{file=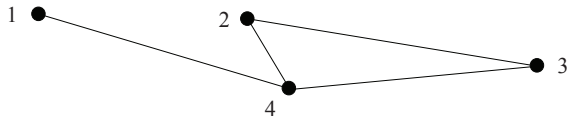,scale=0.35}}
\vspace{-4.7cm}
\caption{The graph $G$ with $V=\{1,2,3,4\}$ and
 $E=\{(1,4),(2,3),(2,4),(3,4)\}$.}
\label{figure5}
\end{figure}

Let $G=(V,E)$ be a graph with a set of vertices $V=\{1,\ldots,n\}$ and a
set of edges $E \subseteq V\times V$. The graph state $|G\rangle$ that
corresponds to the graph $G$ is the quantum state obtained by the
following procedure, where we assume that we have the initial state
$|0\rangle_1\cdots|0\rangle_n$ and call the $k$-th qubit the qubit
corresponding to the vertex $k$:
\begin{enumerate}
\item Apply $H$ to the qubit corresponding to the vertex $k$ for any $k
      \in V$.

\item Apply $\Lambda Z$ to the pair of qubits corresponding to the
      vertices $k_1$ and $k_2$ for any $(k_1,k_2) \in E$.
\end{enumerate}
We call this procedure the standard procedure. For example, the graph
state $|G\rangle$ corresponding to the graph $G$ depicted in
Fig. \ref{figure5} is obtained by
$$\Lambda Z_{14}\Lambda Z_{23}\Lambda Z_{24}\Lambda
Z_{34}H_1H_2H_3H_4|0\rangle_1|0\rangle_2|0\rangle_3|0\rangle_4.$$ 

We consider a quantum circuit for preparing graph states, where the
circuit consists only of projective measurements. As in the standard
quantum circuit model, the complexity measures of a quantum circuit are
the number of qubits in it and its size and depth \cite{Takahashi}. The
size is the number of projective measurements and the depth is the
number of layers in the circuit, where a layer consists of projective
measurements that can be performed simultaneously. A quantum circuit can
use ancillary qubits, which start in state $|0\rangle$.

We show that the set of observables ${\cal S}_3 = \{Z\otimes X,Y\}$ with
one ancillary qubit is sufficient for preparing graph states
efficiently. From the proof of Theorem 1, we can simulate $H$ and
$(P^{-1}\otimes HP^{-1})\Lambda Z (I\otimes H)$ using projective
measurements described by the observables in ${\cal S}_3$. Thus, we can
simulate $(P^{-1}\otimes P^{-1})\Lambda Z$. However, since ${\cal S}_3$
has only $Z\otimes X$ and $Y$, it is difficult to simulate $P$ and thus
$\Lambda Z$. This means that it is difficult to use the standard
procedure directly.

Our circuit consists of three steps. In Step 2, we use $(P^{-1}\otimes
P^{-1})\Lambda Z$ in place of $\Lambda Z$ in Step 2 of the standard
procedure. Since $P^{-1}$ and $\Lambda Z$ commute, this step is
equivalent to Step 2 of the standard procedure up to local unitary gates
generated by $P^{-1}$. We need to remove the side effects, that is, the
local unitary gates, to obtain an exact graph state. If the degree of
the vertex $k$ (that is, the number of edges incident to $k$) is even,
$P^{-1}$ is applied to the qubit corresponding to the vertex $k$ even
times and thus the local unitary gate is $(P^{-1})^2=\sigma_z$ or
$(P^{-1})^4=I$. Similarly, if the degree is odd, the local unitary gate
is $P^{-1}$ or $(P^{-1})^3=\sigma_zP^{-1}$.

\begin{figure}[t]
\centerline{\psfig{file=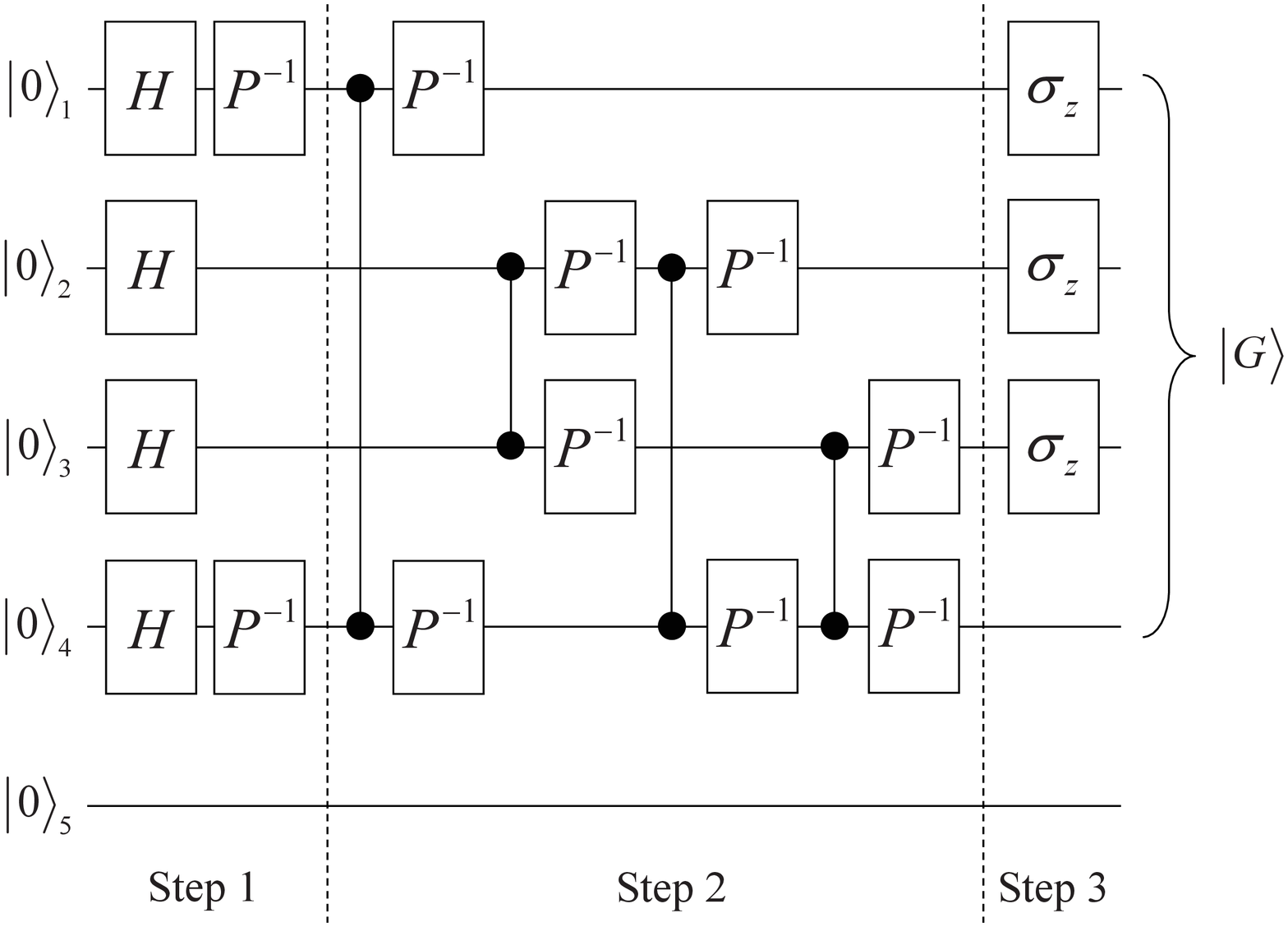,scale=0.3}}
\vspace{-.1cm}
\caption{The quantum circuit for preparing $|G\rangle$ corresponding to
 the graph $G$ depicted in Fig. \ref{figure5}. The gate represented by
 two solid circles connected by a line is $\Lambda Z$. An ancillary
 qubit is reused to simulate $(P^{-1}\otimes P^{-1})\Lambda Z$ and
 $\sigma_z$.}
\label{figure6}
\end{figure}

Our idea of removing the side effects is that, in Step 1 of our circuit,
we apply $H$ (or $\sigma_zH$) to the qubit corresponding to a vertex if
the degree of the vertex is even and we perform a $Y$-measurement on the
qubit to prepare $|-_{\frac{\pi}{2}}\rangle=P^{-1}H|0\rangle$ (or
$|+_{\frac{\pi}{2}}\rangle=\sigma_zP^{-1}H|0\rangle$) if the degree is
odd. Combining Step 1 with Step 2 transforms the side effects in Step 2
to only $\sigma_z$ or $I$. In Step 3, $\sigma_z$ is removed if
needed. For example, our circuit for preparing the graph state
$|G\rangle$ corresponding to the graph $G$ depicted in
Fig. \ref{figure5} is based on the circuit (in the standard quantum
circuit model) depicted in Fig. \ref{figure6}. An ancillary qubit is
reused to simulate $(P^{-1}\otimes P^{-1})\Lambda Z$ and $\sigma_z$.

On the basis of the idea, we show the following theorem, where we assume
that we have a classical description of a given graph and can thus use
the degree of a vertex to construct a quantum circuit:
\begin{theorem}
For any graph $G=(V,E)$, the graph state $|G\rangle$ can be prepared by
 a quantum circuit consisting of one-qubit and two-qubit projective
 measurements described by the observables in ${\cal S}_3 = \{Z\otimes
 X,Y\}$ with size and depth $O(n+m)$ and one ancillary qubit, where
 $n=|V|$ and $m=|E|$.
\end{theorem}
{\it Proof.}
Let $\deg(k)$ be the degree of the vertex $k$. We assume that we have
 the initial state $|0\rangle_1\cdots|0\rangle_{n+1}$ and the $(n+1)$-th
 qubit is an ancillary qubit. Our circuit is constructed by using the
 following procedure:
\begin{enumerate}
\item For $k=1,\ldots,n$:
\begin{itemize}
\item If $\deg(k)\equiv 0 \mod 2$, apply the simulation step of
      $H$ where the qubit corresponding to the vertex $k$ is used as an
      ancillary qubit and the qubit corresponding to the vertex $k+1$
      (in state $|0\rangle$) is used as an input qubit. Let $t_k$ be the
      classical outcome of the $Z\otimes X$-measurement in the
      simulation step. The resulting state of the qubit corresponding to
      the vertex $k$ is $H|0\rangle$ if $t_k=1$ and $\sigma_zH|0\rangle$
      otherwise.

\item If $\deg(k)\equiv 1 \mod 2$, perform a $Y$-measurement on
      the qubit corresponding to the vertex $k$. Let $u_k$ be the
      classical outcome of the measurement. The resulting state of the
      qubit is $|+_{\frac{\pi}{2}}\rangle=\sigma_zP^{-1}H|0\rangle$ if
      $u_k=1$ and $|-_{\frac{\pi}{2}}\rangle=P^{-1}H|0\rangle$
      otherwise.
\end{itemize}

\item Apply $(P^{-1}\otimes P^{-1})\Lambda Z$ as in Step 2 of the
      standard procedure, where we reuse an ancillary qubit.

\item For $k=1,\ldots,n$:

If one of the following conditions holds, apply $\sigma_z$ to the qubit
      corresponding to the vertex $k$, where we reuse an ancillary
      qubit:
\begin{itemize}
\item $\deg(k)\equiv 0 \mod 4$ and $t_k=-1$.

\item $\deg(k)\equiv 1 \mod 4$ and $u_k=-1$.

\item $\deg(k)\equiv 2 \mod 4$ and $t_k=1$.

\item $\deg(k)\equiv 3 \mod 4$ and $u_k=1$.
\end{itemize}
\end{enumerate}
From the proof of Theorem 1, the above procedure can be done by using
 projective measurements described by the observables in ${\cal
 S}_3$. It is easy to show that the circuit works correctly and that the
 size and depth are $O(n+m)$ and the circuit uses only one ancillary
 qubit.\hfill $\Box$\\

The depth of our circuit is larger than H\o yer et al.'s one. Since the
graph states corresponding to connected graphs seem to be particularly
useful in quantum information processing, we are interested in such
graphs. For a connected graph, $m=\Omega(n)$ and thus the depth of our
circuit is $O(m)$ in this case, which is asymptotically the same as H\o
yer et al.'s one.

\section{Conclusions and Future Work}

We showed that the set of observables $\{Z\otimes X,(\cos\theta) X +
(\sin\theta) Y \ {\rm all} \ \theta\in [0,2\pi)\}$ with one ancillary
qubit is universal. This improves Jorrand and Perdrix's result and the
proof immediately implies the best known result for the approximate
universality by Perdrix. The proof also implies that the set of
observables $\{Z\otimes X,Y\}$ with one ancillary qubit is sufficient
for preparing graph states efficiently. It would be interesting to
investigate whether our result can be improved or not. For example, is
there a set of one one-qubit observable and one two-qubit observable
that is approximately universal for quantum computation using one
ancillary qubit?

\section*{Acknowledgments}

The author thanks Yasuhito Kawano, Seiichiro Tani, and Go Kato for their
helpful comments.

\end{document}